\documentclass[aps,prl,superscriptaddress]{revtex4}

\usepackage{graphicx}

\usepackage{verbatim}
\usepackage{graphicx}

\normalfont

\begin{document}


\title{Optical investigation of the natural electron doping in thin MoS$_2$ films deposited on dielectric substrates}

\author{D. Sercombe$^1$, S. Schwarz$^1$, O. Del Pozo-Zamudio$^1$, F. Liu$^{1,2}$, B. J. Robinson$^3$, E. A. Chekhovich$^1$, I. I. Tartakovskii$^4$, O. Kolosov$^3$, A. I. Tartakovskii$^1$\\
{\it $^{1}$Department of Physics and Astronomy, University of Sheffield, Sheffield S3 7RH, United Kingdom\\
$^{2}$Experimentelle Physik 2, Technische Universit\"{a}t Dortmund, 44221 Dortmund, Germany\\
$^3$Department of Physics, University of Lancaster, Lancaster LA1 4YB, United Kingdom\\
$^4$ Institute of Solid State Physics, Russian Academy of Sciences, Chernogolovka, 142432, Russia}}

\begin{abstract}
\end{abstract}

\date{\today}

\maketitle

{\bf Two-dimensional (2D) compounds provide unique building blocks for novel layered devices and hybrid photonic structures. However, large surface-to-volume ratio in thin films enhances the significance of surface interactions and charging effects requiring new understanding. Here we use micro-photoluminescence (PL) and ultrasonic force microscopy to explore the influence of the dielectric environment on optical properties of a few monolayer MoS$_2$ films. PL spectra for MoS$_2$ films deposited on SiO$_2$ substrates are found to vary widely. This film-to-film variation is suppressed by additional capping of MoS$_2$ with SiO$_2$ and Si$_x$N$_y$, improving mechanical coupling of MoS$_2$ with surrounding dielectrics. We show that the observed PL non-uniformities are related to strong variation in the local electron charging of MoS$_2$ films. In completely encapsulated films, negative charging is enhanced leading to uniform optical properties. Observed great sensitivity of optical characteristics of 2D films to surface interactions has important implications for optoelectronics applications of layered materials.}\\



Interest in atomically thin two-dimensional (2D) layered compounds is growing due to unique physical properties found for monolayer (ML) structures \cite{NovoselovPNAS2005,WangNatNano2012}. One such material, molybdenum disulfide (MoS$_{2}$), has generated particular interest due to the presence of an indirect-to-direct band gap transition and observation of photoluminescence (PL)\cite{SplendianiNanoLett2010,MakPRL2010,EdaNanoLett2011} and electro-luminescence \cite{SundaramNanoLett2013} in the visible range up to room temperature. A high on/off ratio (exceeding 10$^{8}$) has suggested a potential use in field effect transistors \cite{RadisavljevicNatNano2011}, while a strong valley polarization is likely to be used in the development of future valleytronics applications \cite{MakNatNano2012,SallenPRB2012,ZengNatNano2012,CaoNatComm2012,XiaoPRL2012}.  

PL studies \cite{MakPRL2010} have shown that suspended ML films of MoS$_{2}$ have enhanced emission compared to those in contact with the substrate. However, the eventual integration of MoS$_{2}$ into devices, such as transistors and photonic structures, will render the use of suspended MoS$_{2}$ impractical. The effect of dielectric encapsulation has so far been reported for high-$k$ dielectric materials commonly used in transistors \cite{JenaPRL2007,PlechingerPSS2012,YanArxiv2012}. However, no thorough study showing how dielectric environments affect optical properties and charging of MoS$_{2}$ films has yet been conducted, the two issues we address in this work. 

We focus on interaction of MoS$_2$ films with SiO$_{2}$ and Si$_x$N$_{y}$ commonly used in photonic devices, and report low temperature PL measurements on over a 100 thin films, enabling detailed insight in interactions of MoS$_{2}$ with its dielectric surrounding. We study mechanically exfoliated MoS$_{2}$ films deposited on silicon substrates finished with either nearly atomically flat thermally grown SiO$_2$ or relatively rough SiO$_2$ grown by plasma-enhanced chemical vapor deposition (PECVD). We use a combination of low temperature micro-photoluminescence (PL), atomic force microscopy (AFM) and ultrasonic force microscopy (UFM). We find marked variety of the PL spectral lineshapes and peak energies in the large number of few monolayer MoS$_{2}$ films, which nonetheless show trends that we are able to relate to electrostatic and mechanical interaction of thin films with the surrounding dielectrics. 

We find that high mechanical coupling between a MoS$_{2}$ film and the surrounding layers is only possible for capped films on thermally grown SiO$_2$, whereas more complex morphology and poorer contact with the surrounding layers is observed for uncapped films, the effect further exacerbated for films on PECVD substrates. Following this observation, we show a direct correlation between the enhanced mechanical coupling of MoS$_{2}$ with the surrounding dielectric layers and increased negative charging of the films, directly affecting spectral characteristics of PL due to the presence of the pronounced PL peak of a negatively charged trion. By comparison with recent work where MoS$_{2}$ films were controllably charged by applying voltage \cite{MakNatMat2012}, we estimate that observed charging lead to electron densities of the order of $10^{12}$ cm$^{-2}$.  
 
\section{Results}

\subsection{PL of thin MoS$_2$ films}

We carry out low-temperature PL spectroscopy on a wide range of MoS$_2$ films having thicknesses between 2 and 5 ML, as estimated from the careful colour-contrast examination using optical microscopy, and further confirmed for some films from AFM measurement (see Methods for description of sample fabrication and PL techniques). Fig.1(a) shows a selection of PL spectra measured for uncapped MoS$_2$ films deposited on Si substrates with either PECVD (a-d) or thermal oxidation (e-h). In all spectra exciton complexes $A$ and $B$ are clearly visible \cite{SplendianiNanoLett2010}, although there is a large variation in PL lineshapes for different films. The $A$ complex is composed of a trion PL peak $A^-$ and a high energy shoulder $A^0$ corresponding to neutral exciton PL \cite{MakNatMat2012}. A low energy shoulder $L$ is also observed in some spectra, though spectra showing weak or no contributions of $L$ and $A^0$ states were observed on both PECVD (a) and thermal oxide (e) substrates. A relatively large contribution of $L$ and $A^0$ was found in many films deposited on PECVD substrates (b, c) and in some cases the neutral exciton was found to have brighter emission than the trion [as in (d)]. For films deposited on thermal oxide substrates, there is a less significant variation in the lineshape (e-h), and  $L$ and $A^0$ features are, in general, less pronounced relative to $A^-$ than in films deposited on PECVD grown SiO$_2$.

The effect of additional capping of MoS$_2$ films with dielectric layers is demonstrated in Fig.2. A 100 nm thick layer of either SiO$_{2}$ or Si$_x$N$_{y}$ is deposited using PECVD on top of the MoS$_2$/SiO$_{2}$/Si samples for both PECVD and thermal SiO$_2$/Si substrates. Here we observe even less variation in lineshapes between the films. A further suppression of the low energy shoulder $L$ and neutral exciton peak $A^0$ is found for films capped with Si$_x$N$_{y}$ (a,b,e,f) on both types of substrates, and with SiO$_{2}$ on thermally grown substrates. In contrast, $L$ and $A^0$ peaks are pronounced when capping with SiO$_{2}$ is used for MoS$_2$ films on PECVD substrates. Further to this, from comparison of spectra in (a,b,c,d) and (e,f,g,h), we find that the PL linewidths of films deposited on the PECVD oxide are notably broader than for those on the thermal oxide substrates. 

An interesting trend in all spectra presented in Figs.1 and 2 is a correlation between the intensities of the features $L$ and $A^0$: the two peaks are either both rather pronounced or suppressed in any given spectrum relative to the trion peak $A^-$. This may imply that peak $L$ becomes suppressed when the film captures an excess of negative charge.

\subsection{Analysis of PL peak energies}

A statistical analysis of PL peak energies for films deposited on the two types of substrates is presented in Fig.3. Fig.3(a,b) show that the average values for the PL peak energies, $E^{av}_{max}$, for uncapped films are $E^{av}_{max}=1.888$ eV for the PECVD substrates and $E^{av}_{max}=1.880$ eV for thermal oxide substrates, with an almost two times larger standard deviation, $\sigma_{Emax}$ for the former (18 versus 11 meV). The data collected for the capped films (shaded for Si$_{x}$N$_{y}$ and hatched for SiO$_{2}$) are presented in Fig.3(c) and (d) for the thermal and PECVD oxide substrates, respectively. Significant narrowing of the peak energy distribution is found in all cases: $\sigma_{Emax} \approx$6 meV has been found. The average peak energies are very similar for both  SiO$_2$ and Si$_{x}$N$_{y}$ capping on the thermal oxide substrates ($E^{av}_{max}$=1.874 eV), but differ for PECVD substrates: $E^{av}_{max}$=1.862 and 1.870 eV for SiO$_2$ and Si$_{x}$N$_{y}$ capping, respectively. 

From previous reports \cite{MakPRL2010}, for films with thicknesses in the range 2 to 5 MLs, one can expect the PL peak shift on the order of 20 meV. In addition, PL yield was reported to be about 10 times higher for 2 ML films compared with 4 ML and for 3 ML compared with 5 ML \cite{MakPRL2010}. In our study, the integrated PL signal shows a large variation within about one order of magnitude between the films. The dependence of the PL yield on the type of the substrate and capping is not very pronounced. While our data for PL intensities is consistent with the reported in the literature for the range of thicknesses which we studied, the PL peak energy distribution shows the unexpected broadening for uncapped samples: for example, deviations from $E^{av}_{max}$ by $\pm$20-30 meV are evident in Fig.3(a,b). For the capped samples, new trends are observed: the significant narrowing and red-shift of $E_{max}$ distributions. As shown below, these effects reflect changes in the PL lineshapes between the capped and uncapped samples, which in their turn reflect changes in the relative intensities of the $A^-$, $A^0$ and $L$ peaks. 

We note that the new experimental trends observed in our PL studies do not depend on the exact distribution of thicknesses in the ensembles of the investigated films, provided these distributions are similar for all types of samples studied. The latter is the case in our study, as the films were produced using the same method, show similar range of the colour-contrasts under optical examination, and exhibit similar ranges of PL yield.    

\subsection{Analysis of PL lineshapes}

In this section  we will present the lineshape analysis for the $A$ exciton PL based on the measurement of full width at half maximum (FWHM) in each PL spectrum. This approach allows one to account for contributions of the three PL features, $L$, $A^0$ and $A^-$. The data are summarized in Fig.4 and Tables \ref{Tab1}.

{\bf PECVD grown SiO$_2$ substrates.} These data are presented in Fig.4 in red. Data for uncapped films are shown in Fig.4(a), from where it is evident that the lineshapes vary dramatically from film to film within a range from 50 to 170 meV. FWHM for uncapped films on PECVD grown substrates is on average $\Delta E^{av}_{FWHM}$=96 with a large standard deviation $\sigma_{FWHM}$=33 meV. This gives a rather high coefficient of variation $\sigma_{FWHM}/\Delta E^{av}_{FWHM}$=0.34 showing normalized dispersion of the distribution of the PL FWHM.

\begin{table}[t]
\caption{\label{Tab1} Mean values, standard deviations and coefficients of variation for full width at half maximum of PL spectra measured for thin MoS$_{2}$ films.}  
\centering  
\begin{tabular}{l c c c}  
\hline\hline                       
  Substrate/Capping  & Mean value & Standard deviation  & Coefficient of variation  \\[1ex]
\hline              
{PECVD/uncapped} & 96 meV & 33 meV & 0.34\\[1ex]
{PECVD/SiO$_2$} & 109 meV & 9 meV & 0.08\\[1ex]
{PECVD/Si$_x$N$_y$} & 84 meV & 7 meV & 0.08\\[1ex]
{Thermal/uncapped} & 79 meV & 12 meV & 0.15\\[1ex]
{Thermal/SiO$_2$} & 76 meV & 7 meV & 0.09\\[1ex]
{Thermal/Si$_x$N$_y$} & 64 meV & 4 meV & 0.06\\[1ex]
\hline                          
\end{tabular}
\end{table}

The non-uniformity of lineshapes of the PL spectra is significantly suppressed by capping the films with Si$_x$N$_{y}$ and SiO$_2$ (shown with red in Fig.4(b) and (c), respectively). This is evidenced from the reduction of the coefficient of variation in the FWHM values  by a factor of 4 in capped films compared with the uncapped samples (in Table \ref{Tab1}). Despite the narrowed spread of $\Delta E_{FWHM}$ values, the average FWHM in SiO$_2$ capped films is rather high, 109 meV, which reflects relatively strong contribution of $L$ and $A^0$ PL features. Contributions of $A^-$, $L$ and $A^0$ features vary very considerably in the uncapped samples, leading to on average smaller linewidths but a very significant spread in FWHM values. In contrast, in  Si$_x$N$_{y}$ capped films, $A^-$ peak dominates and both $L$ and $A^0$ features are relatively weak, which effectively results in narrowing of PL.

{\bf Thermally grown SiO$_2$ substrates.} These data are presented in Fig.4 in blue. It can be seen that uncapped films deposited on the flatter thermal oxide substrates appear to have significantly narrower distributions of linewidths compared to uncapped films on PECVD substrates: coefficient of variation of $\Delta E_{FWHM}$ is by a factor of 2 smaller for films on the thermally grown substrates [see Fig.4(a) and Table \ref{Tab1}]. In addition, compared with the films deposited on PECVD grown SiO$_2$, FWHM is also reduced by about 20$\%$ to 79 meV. Such narrowing reflects weaker contribution of $L$ and $A^0$ peaks in PL spectra.

The non-uniformity of the PL spectra still present in uncapped films deposited on thermally grown SiO$_2$ is further suppressed by capping the films with Si$_x$N$_{y}$ and SiO$_2$ [shown with blue in Fig.4(b) and (c), respectively]. In general, the coefficients of variation for FWHM of the capped films are rather similar for both substrates and are in the range of 0.06-0.09, showing significant improvement of the reproducibility of PL features compared with the uncapped samples (see Table \ref{Tab1}). For Si$_x$N$_{y}$ capped films on thermally grown SiO$_2$, we also observe narrowing of PL emission to $\Delta E^{av}_{FWHM}$=64 meV. This reflects further suppression of $L$ and $A^0$ peaks relative to $A^-$, the effect less pronounced in SiO$_2$ capped films.

\subsection{AFM and UFM measurements}

To further understand the interactions between MoS$_2$ films and the substrate/capping materials, we carried out detailed AFM and UFM measurements of our samples (Fig.5). AFM measurements of films deposited on PECVD grown substrates Fig.5(a) show that the film is distorted in shape and follows the morphology of the underlying substrate. The root mean square (rms) roughness $R_{rms}$ of these films is 1.7 nm with a maximum height $R_{max}$=11 nm, similar to the parameters of the substrate, $R_{rms}$=2 nm and $R_{max}$=15 nm. Such $R_{max}$ is greater than the thickness of films ($<$3 nm), leading to significant film distortions. UFM measurements of these films [Fig.5(b)] show small areas of higher stiffness (light colour, marked with arrows) and much larger areas of low stiffness (i.e. no contact with the substrate) shown with a dark colour. This shows that the film is largely suspended above the substrate on point contacts. 

AFM measurements of films deposited on thermally grown SiO$_2$ substrates [Fig.5(c)] show a much more uniform film surface due to the less rough underlying substrate. This is reflected in a significantly improved $R_{rms}$ = 0.3 nm and $R_{max}$=1.8 nm. These values are still higher than those for the bare substrate with $R_{rms}$ = 0.09 nm and $R_{max}$=0.68 nm. A more uniform stiffness distribution is observed for these films in UFM [Fig.5(d)], although the darker colour of the film demonstrates that it is much softer than the surrounding substrate and thus still has relatively poor contact with the substrate. A darker shading at film edges demonstrates that they have poorer contact than the film center and effectively curl away from the substrate. 

AFM and UFM data for films capped with 15 nm SiO$_{2}$ after deposition on PECVD and thermally grown SiO$_2$ are given in Fig.5(e, f) and (g, h),  respectively. For the PECVD substrate, the roughness of the MoS$_2$ film is similar to that in the uncapped sample in Fig.5(a): $R_{rms}$=1.68 nm and $R_{max}$=10.2 nm. From the UFM data in Fig.5(f), it is evident that although the contact of the MoS$_2$ film with the surrounding SiO$_2$ is greatly improved compared with the uncapped films, a large degree of non-uniformity is still present, as concluded from many dark spots on the UFM image. In great contrast to that, the capped MoS$_2$ film on thermally grown SiO$_2$ is flatter [Fig.5(g, h)], $R_{rms}$=0.42 nm and $R_{max}$=6.1 nm, with the roughness most likely originating from the PECVD grown SiO$_2$ capping layer. The UFM image in Fig.5(h) shows remarkable uniformity of the stiffness of the film similar to that of the capped substrate, demonstrating uniform and firm contact (i.e. improved mechanical coupling) between the MoS$_2$ film and the surrounding dielectrics.

\section{Discussion}

There is a marked correlation between the PL properties of the MoS$_2$ films and film stiffness measured by UFM. The stiffness reflects the strength of the mechanical coupling between the adjacent monolayers of the MoS$_2$ film and the surrounding dielectrics. The increased bonding and its uniformity for films deposited on less rough thermally grown SiO$_2$ substrates and for capped MoS$_2$ films manifests in the more reproducible PL characteristics, leading to reduced standard deviations of the peak positions and linewidths. These spectral characteristics are influenced by the relative intensities of the three dominating PL features, trion $A^-$, neutral exciton $A^0$ and low energy $L$ peak, which are influenced by the charge balance in the MoS$_2$ films sensitive to the dielectric environment. The efficiency of charging can be qualitatively estimated from the relative intensities of $A^-$ and $A^0$ peaks. In the vast majority of the films, $A^-$ dominates. As noted above, the intensity of $A^0$ directly correlates (qualitatively) with that of the relatively broad low energy PL shoulder $L$ (see Fig.1 and 2), previously ascribed to emission from surface states. The lineshape analysis presented in Fig.4 and Table \ref{Tab1} is  particularly sensitive to the contribution of peak $L$. 

The PL lineshape analysis and comparison with the UFM data lead to conclusion that negative charging of the MoS$_2$ films is relatively inefficient for partly suspended uncapped films on rough PECVD substrates. Both in SiO$_2$ and Si$_x$N$_{y}$ capped films on PECVD substrates, the charging effects are more pronounced. However, both $A^0$ and $L$ features still have rather high intensities. The relatively low charging efficiency is most likely related to a non-uniform bonding between the MoS$_2$ films and the surrounding dielectric layers as concluded from from UFM data [see Fig.5(f)]. The charging is more pronounced for uncapped MoS$_2$ films on thermal oxide substrates, and is enhanced significantly more for capped films: for Si$_x$N$_{y}$ capping $A^0$ and $L$ peaks only appear as weak shoulders in PL spectra. 

It is clear from this analysis that the charge balance in the MoS$_2$ films is altered strongly when the films are brought in close and uniform contact with the surrounding dielectrics, enabling efficient transfer of charge in a monolithic hybrid heterostructure. Both $n$-type\cite{RadisavljevicNatNano2011,MakPRL2010,DoluiPRB2013} and $p$-type \cite{ZhanSmall2012,DoluiPRB2013} conductivities have been reported in thin MoS$_2$ films deposited on SiO$_2$. It is thus possible that the sign and density of charges in exfoliated MoS$_2$ films may be strongly affected by the properties of PECVD grown SiO$_2$ and Si$_x$N$_{y}$, where the electronic properties may vary depending on the growth conditions \cite{WolfJAP2005,Boogaard2009,Zou2011}. It is notable, however, that for a large variety of samples studied in this work, the negative charge accumulation in the MoS$_2$ films is pronounced, and is further enhanced when the bonding of the films with the dielectric layers is improved.

In order to estimate the density of the accumulated charges we refer to Ref.\cite{MakNatMat2012}, where PL spectra as a function of electron density in the film were measured. The neutral exciton PL peak $A^0$ becomes less intense than the trion peak $A^-$ at the electron density $n\approx 2\times 10^{12}$ cm$^{-2}$. Since in our experiments in many films $A^0$ peak is relatively pronounced, we conclude that we have studied the regime where the electron densities are of the order of $10^{12}$ cm$^{-2}$ or less.

The band-structure of MoS$_2$ and hence its optical characteristics can also be influenced by strain \cite{PeelaersPRB2012,ConleyArxiv2013,HeArxiv2013}. However, the distribution and magnitude of strain cannot be assessed directly in our experiments. Indirect evidence for increased tensile strain in capped samples compared to uncapped films on PECVD may be deduced from the red-shift of the average PL peak energy by $approx$20 meV after capping (data in Fig.3). On the other hand, doping-dependent Stokes shifts of the trion PL have been found recently \cite{MakNatMat2012}, which may explain the behavior we find in charged MoS$_2$ sheets. One would expect a more uniform strain distribution in the case of uniform mechanical properties of the sample, which as shown by UFM is achieved for capped MoS$_2$ films on flat thermally grown SiO$_2$ substrates. In order to roughly estimate a possible magnitude of strain in our films we refer to recent work in Refs.\cite{HeArxiv2013,CastellanosGomezArxiv2013,WangArxiv2013} reporting 50 to 80 meV bandgap shift per percent of applied strain. This implies that in our experiments the maximum strain in the films reaches $\approx 0.5\%$. 

We note, that another method for sensing the strain and charging in thin MoS$_2$ films is by using Raman spectroscopy. Recently it has been found that Raman modes \cite{ZhangPRB2013} $E^1_{2g}$ and $A_{1g}$ exhibit frequency shifts under the influence of strain \cite{RicePRB2013,CastellanosGomezArxiv2013,WangArxiv2013} and charging  \cite{ChakrabortyPRB2012}. We have also carried out preliminary Raman experiments on a set of films placed on PECVD and thermally grown SiO$_2$ substrates and capped with SiO$_2$ and Si$_x$N$_y$. However, only very weak trends have been observed. This is consistent with relatively low charging densities and strain in our samples, the effect of which is readily detectable in PL \cite{MakNatMat2012,CastellanosGomezArxiv2013,WangArxiv2013}, but is less pronounced in Raman measurements.

In conclusion, we demonstrate that it is possible to increase the reproducibility of optical characteristics of mechanically exfoliated few mono-layer MoS$_{2}$ films by coating the films with additional dielectric layers of either SiO$_2$ or Si$_x$N$_{y}$. By comparing PL data with results obtained in UFM, we show that there is a direct correlation between the degree of the mechanical coupling of the MoS$_{2}$ films to the surrounding dielectrics and uniformity of the optical properties. We show that a wide spread in PL spectral lineshapes occurs in general as a result of the film-to-film variation of the relative intensities of the negatively charged trion peak $A^-$ and the two other features, neutral exciton peak $A^0$ and a low energy PL band $L$. We find that when the mechanical coupling between the films and the dielectrics is improved, the films become increasingly negatively charged, as deduced from the pronounced increase in PL of the trion peak, dominating in the majority of PL spectra. Such charging, and also possibly reduction in strain non-uniformities, underpins the highly uniform PL properties in capped MoS$_2$ films, leading to the smallest linewidths below $70$ meV for thin MoS$_2$ films deposited on thermally grown SiO$_2$ and capped with a Si$_x$N$_{y}$ layer.    

\section{Methods}

{\bf Sample preparation.} MoS$_{2}$ was exfoliated using the mechanical cleavage method \cite{NovoselovPNAS2005} and deposited on commercially purchased Si wafers with a low roughness 300 nm thick thermally grown SiO$_{2}$ \cite{BenameurNanotech2011}. Further MoS$_{2}$ samples were produced using the same technique, but deposited on Si substrates covered with 300 nm PECVD grown SiO$_{2}$. PECVD deposition was done in a 60$^{o}$C chamber with a sample temperature of 300$^{o}$C. The root mean square (rms) roughness, $R_{rms}$, of the PECVD grown SiO$_{2}$ is found to be 2 nm with a maximum peak height of 15 nm, whereas $R_{rms}$ of the thermally grown SiO$_{2}$ is 0.09 nm with a maximum height of 0.68 nm. The additional capping of the MoS$_2$/SiO$_2$/Si samples with Si$_x$N$_y$ and SiO$_2$ was carried out using the same PECVD techniques. The complete SiO$_2$/MoS$_2$/SiO$_2$/Si or Si$_x$N$_y$/MoS$_2$/SiO$_2$/Si samples had the top Si$_x$N$_y$ and SiO$_2$ layers with thicknesses of 100 nm for PL and 15 nm for AFM/UFM measurements. The high surface roughness rendered impractical use of AFM for measurements of the film thicknesses on PECVD substrates. However, independent of the substrate type, the uncapped thin MoS$_2$ films had optical contrasts corresponding to thicknesses of 2-5 MLs, further confirmed by AFM on thermal oxide substrates. Furthermore, the range of PL yield was similar for all four types of sample independent of the substrate and capping type, further indicating that ensembles of films with similar distributions of film thicknesses were measured. 

{\bf Micro-photoluminescence experiments.} Low temperature (10K) micro-PL was carried out on a large number of thin films in a continuous flow He cryostat. The signal was collected and analyzed using a single spectrometer and a nitrogen-cooled charged-coupled device. The sample was excited with a laser at 532 nm. All PL spectra presented in this work were measured in a range of low powers where no dependence on power of PL lineshape was found.

{\bf AFM/UFM experiments.} As shown elsewhere \cite{Kolosov1993},the ultrasonic force microscopy (UFM) allows imaging of the near-surface features and subsurface interfaces with superior nanometre scale resolution compared to AFM techniques\cite{YamanakaAPL1994}. In the sample-UFM modality used in this paper \cite{McGuiganAPL2002}, the sample in contact with the AFM tip is vibrated at small amplitude (0.5-2 nm) and high frequency (2-10 MHz), much higher than the resonance frequencies of the AFM cantilever. The resulting sample stress produces a reaction, that is modified by the voids, subsurface defects or sample-substrate interfaces, and can be detected as an additional 'ultrasonic' force. A unique feature of UFM is that it enables nanometre scale resolution imaging of morphology of subsurface nano-structures and interfaces of solid-state objects. In order to interpret the images of a few layer films presented in Fig.5, one can note that the bright (dark) colors correspond to higher (lower) sample stiffness.

\section{Acknowledgements}

This work has been supported by the Marie Curie ITNs S$^3$NANO and Spin-Optronics, EPSRC Programme grant EP/J007544/1, and EU FP7 GRENADA grant. O. D. P.-Z. thanks CONACYT-Mexico Doctoral Scholarship.

\section{Author contributions}

D. S. and S. S. made the samples. D. S., S. S., O. D. P.-Z., F. L., I. I. T. and E. A. C. measured and analyzed optics data. B. J. R. and O. K. measured AFM and UFM. D. S. and A. I. T. wrote the manuscript with input from all co-authors. E. A. C. supervised optical spectroscopy experiments. A. I. T. guided the project.






\newpage

\begin{figure}[t]
\includegraphics[width=0.8\textwidth]{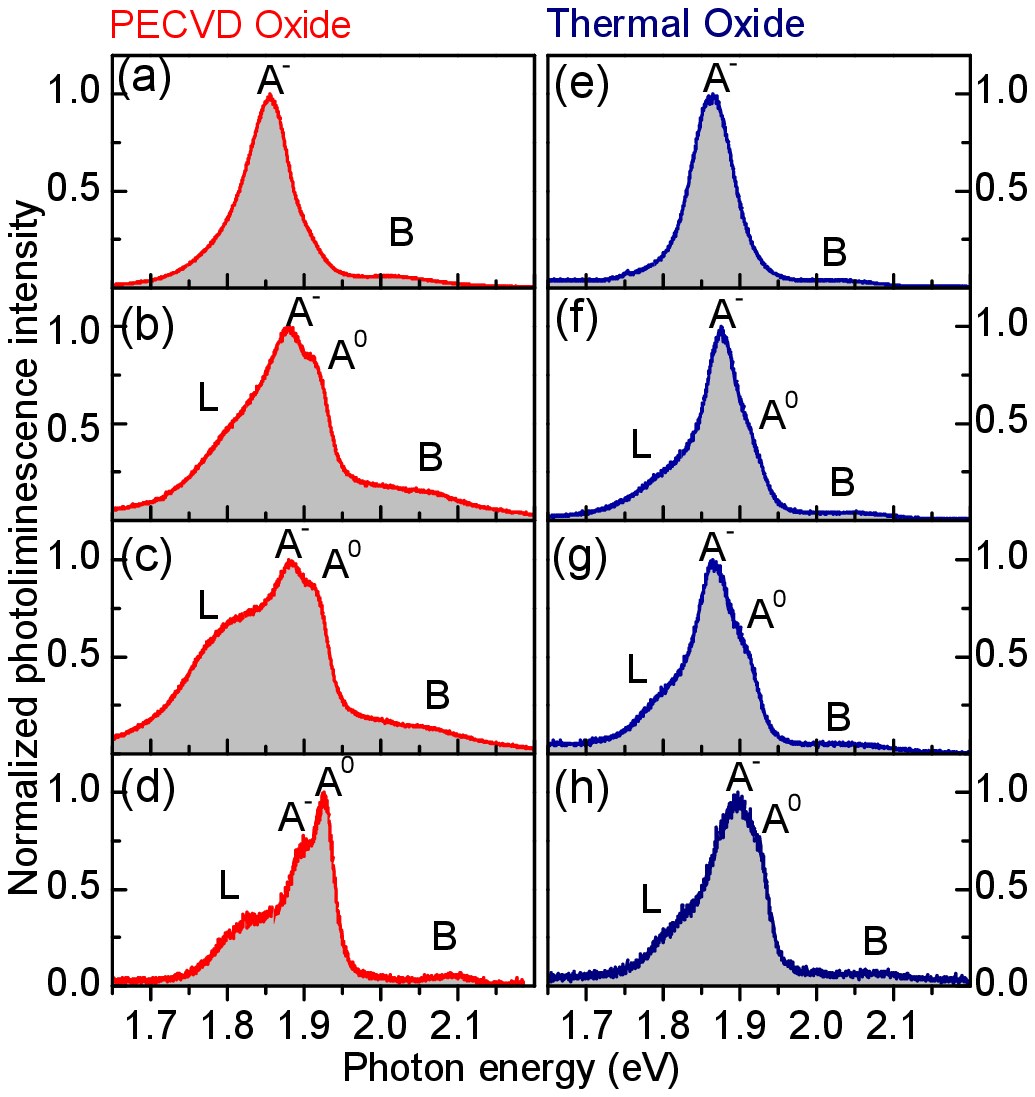}
\caption{PL spectra measured for individual mechanically exfoliated MoS$_{2}$ uncapped films deposited on a 300 nm SiO$_{2}$ layer grown by either PECVD (a-d) or thermal oxidation (e-h) on a silicon substrate.}
\end{figure}

\newpage

\begin{figure}
\includegraphics[width=0.8\textwidth]{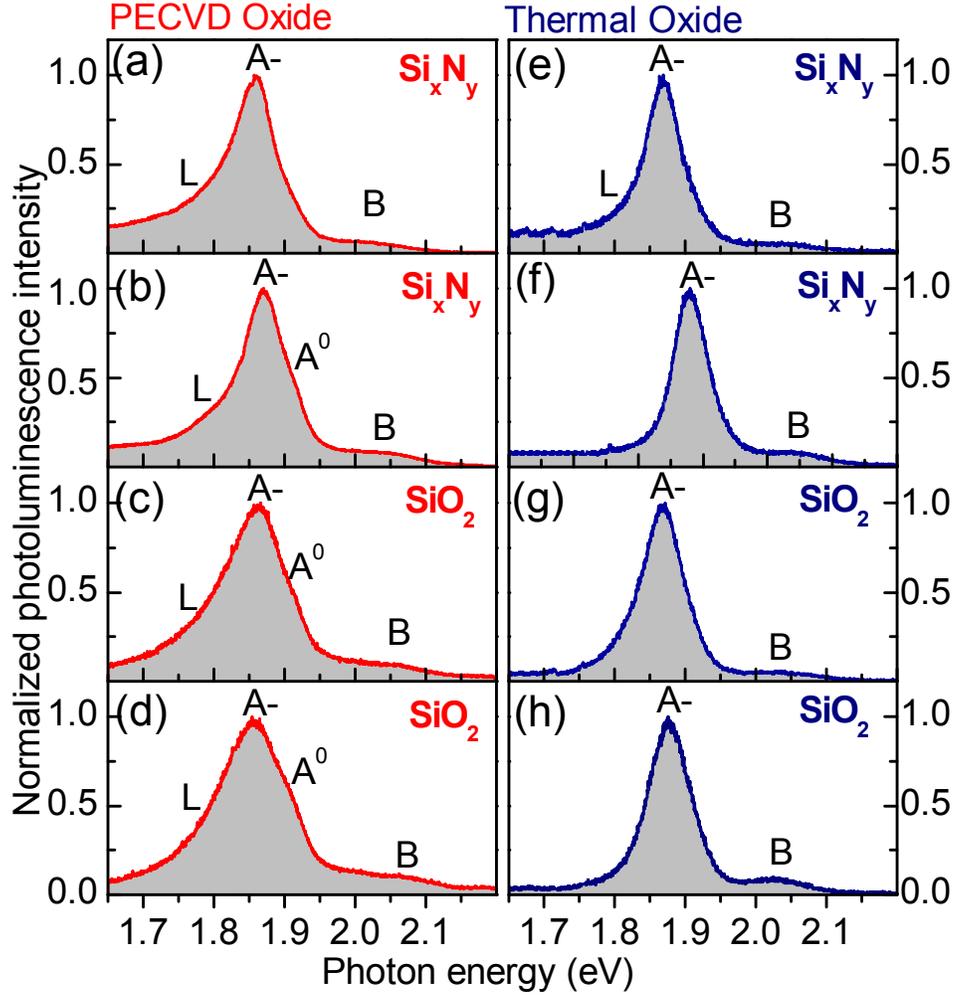}
\caption{PL spectra measured for individual mechanically exfoliated MoS$_{2}$ films capped by a 100 nm PECVD layer of dielectric material. The effect of capping is shown for films deposited on PECVD grown SiO$_{2}$ substrates for SiN (a, b) and SiO$_{2}$ (c, d) capping layers, and also for films deposited on thermally grown SiO$_{2}$ and capped with SiN (e, f) and SiO$_{2}$ (g, h).}
\end{figure}

\newpage

\begin{figure}
\includegraphics[width=0.8\textwidth]{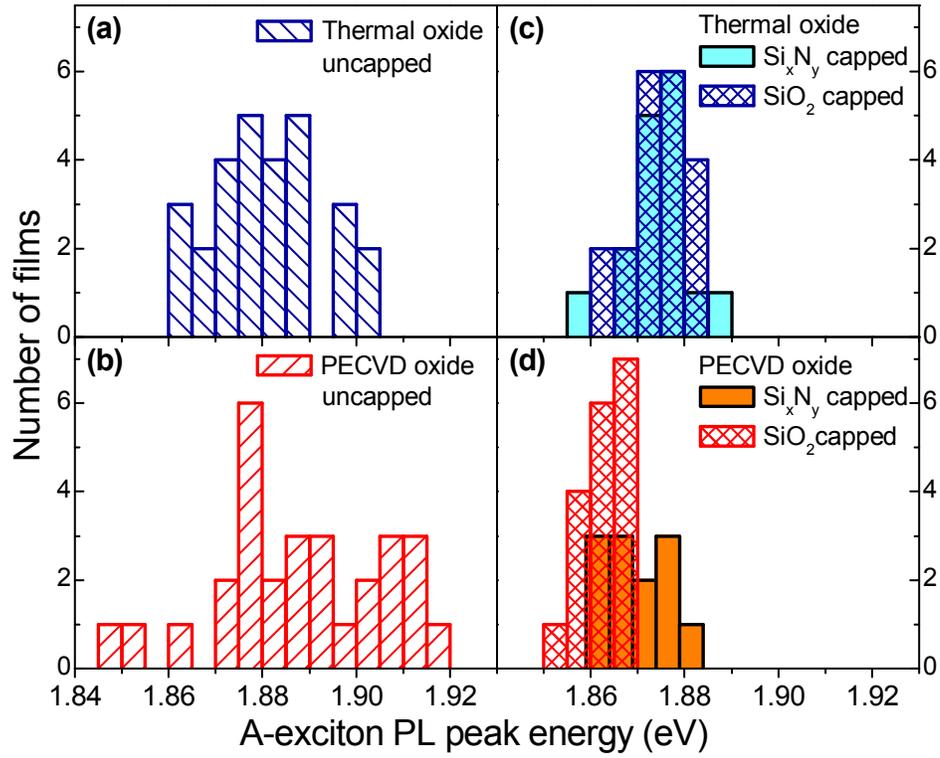}
\caption{(a-d) PL peak energies for $A$ exciton complex in MoS$_2$ thin films. Data for films deposited on thermally (PECVD) grown SiO$_2$ substrates are shown in top (bottom) panels. Panels (a)-(b) and (c)-(d) show PL peak positions for uncapped and capped films, respectively.}
\end{figure}

\newpage

\begin{figure}
\includegraphics[width=0.8\textwidth]{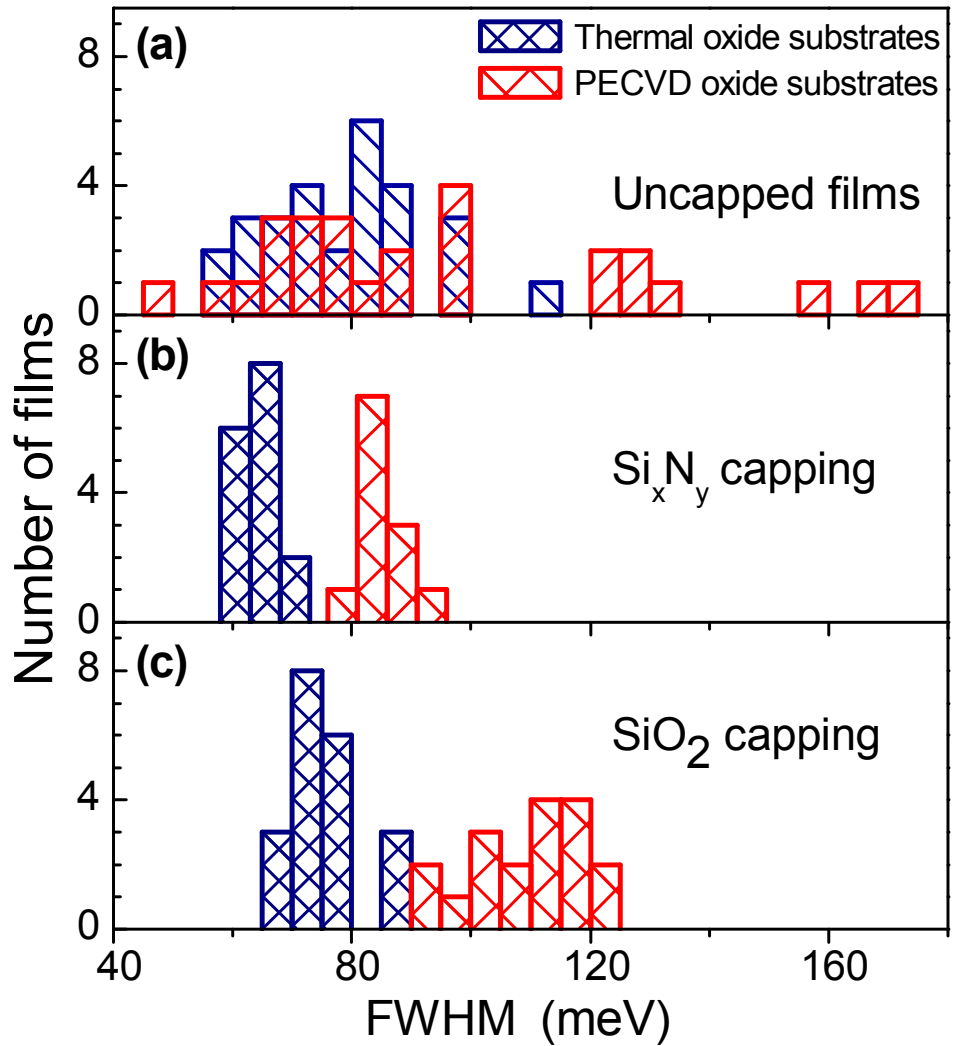}
\caption{PL FWHM of exciton complex $A$ in thin MoS$_2$ films. Data for MoS$_2$ films deposited on thermally and PECVD grown SiO$_2$ substrates is shown with blue and red, respectively. (a) PL FWHM of uncapped MoS$_2$ films. (b) PL FWHM of Si$_{x}$N$_{y}$ capped MoS$_2$ films. (c) PL FWHM of SiO$_2$ capped MoS$_2$ films.}
\end{figure}

\newpage

\begin{figure}
\includegraphics[width=0.8\textwidth] {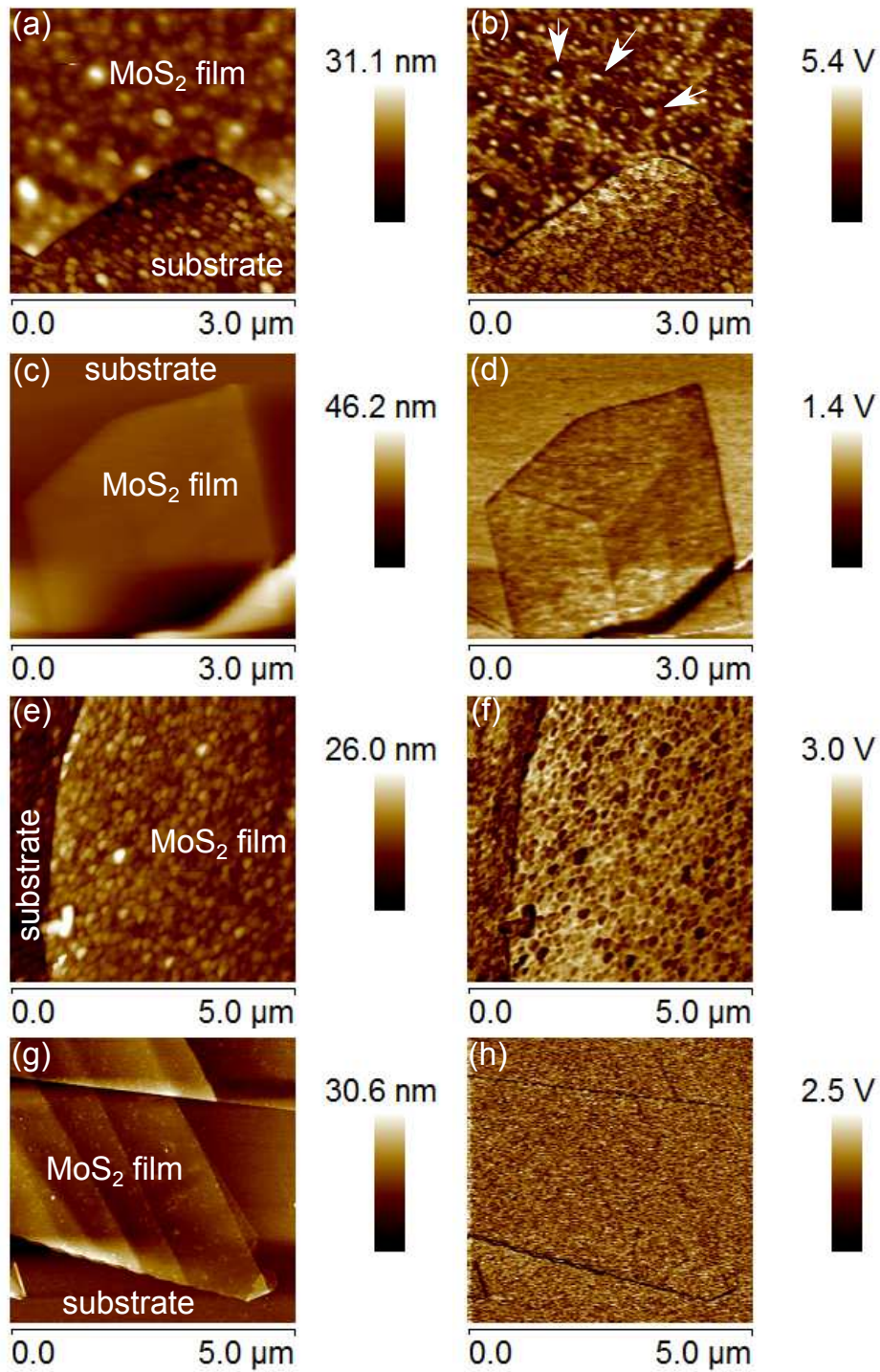}
\caption{AFM (left column) and UFM (right column) images for MoS$_2$ thin films deposited on PECVD and thermally grown SiO$2$ substrates. (a,b) PECVD substrate, uncapped MoS$_2$ film; (c,d) thermally grown substrate, uncapped MoS$_2$ film; (e,f) PECVD substrate, MoS$_2$ film capped with 15 nm of SiO$_2$ grown by PECVD; (g,h) thermally grown substrate, MoS$_2$ film capped with 15 nm of SiO$_2$ grown by PECVD.}
\end{figure}

\end{document}